\documentclass[pre,amsmath,amssymb,showpacs,preprintnumbers]{revtex4}
\usepackage{subfigure}
\usepackage{epsfig}
\usepackage{amsmath}
\usepackage{graphics}
\usepackage{graphicx}
\usepackage{times}
\usepackage{color}

\begin{document}
\title{Phase Diagram and Commensurate-Incommensurate Transitions in the Phase Field
Crystal Model with an External Pinning Potential}
\author{C. V. Achim}
\affiliation{Laboratory of Physics, Helsinki University of
Technology, P.O. Box 1100, FIN-02015 TKK, Finland}
\author{M. Karttunen}
\affiliation{Department of Applied Mathematics, The University of
Western Ontario, London (ON), Canada}
\author{K.R. Elder}
\affiliation{Department of Physics, Oakland University, Rochester,
Michigan, 48309-4487, USA}
\author{E. Granato}
\affiliation{Laborat\'orio Associado de Sensores e Materiais,
Instituto Nacional de Pesquisas Espaciais, S\~ao Jos\'e dos
Campos, SP Brazil}
\author{T. Ala-Nissila}
\affiliation{Laboratory of Physics, Helsinki University of
Technology, P.O. Box 1100, FIN-02015 TKK, Finland and Department
of Physics, Brown University, Providence, R.I. 02912-1843, U.S.A.}
\author{S.C. Ying}
\affiliation{Department of Physics, Brown University, Providence,
R.I. 02912-1843, U.S.A.}

\date{\today}

\begin{abstract}
We study the phase diagram and the commensurate-incommensurate
transitions in a phase field model of a two-dimensional crystal
lattice in the presence of an external pinning potential. The
model allows for both elastic and plastic deformations and
provides a continuum description of lattice systems, such as for
adsorbed atomic layers or two-dimensional vortex lattices.
Analytically, a mode expansion analysis is used to determine the
ground states and the commensurate-incommensurate transitions in
the model as a function of the strength of the pinning potential
and the lattice mismatch parameter. Numerical minimization of the
corresponding free energy shows good agreement with the analytical
predictions and provides details on the topological defects in the
transition region. We find that for small mismatch the transition
is of first-order, and it remains so for the largest values of
mismatch studied here. Our results are consistent with results of
simulations for atomistic models of adsorbed overlayers.
\end{abstract}


\maketitle

%\tableofcontents

\section{Introduction}

In nature there exist many modulated structures which posses two
or more competing length scales. Such systems often exhibit
commensurate-incommensurate (CI) phase transitions
\cite{Schick:81or,Bak:82be}, which are characterized by structural
changes induced by competition between these different scales.
Systems in this class that have received intense attention for
many years include spin-density-waves \cite{Faw01,Ruvalds01} in Cr
and charge-density-wave systems~\cite{Fleming:85ky} \emph{e.g.} in
"blue bronze". They are characterized by an order parameter
(\emph{e.g.}, charge or spin density) that is modulated in space
with a given wave vector $q$. In these systems the transition from
a commensurate (C) state to an incommensurate (I) state is
controlled by temperature and interaction with defects of the
lattice \cite{Millan01}. Other systems of interest are materials
which exhibit magnetic phase with spiral-like structures
\cite{Dzya01,Zhel01}. Finally, vortex lattices in superconducting
films with pinning centers \cite{Martin01} and weakly adsorbed
monolayers~\cite{Thomy:80pl,Pandit:82tu} on a substrate comprise a
2D realization of systems exhibiting CI transitions. In these
systems the interparticle interactions are minimized by a
configuration with a lattice constant $a$, while the
substrate-adsorbed monolayer interaction is minimized by a
configuration with lattice constant $b$ usually incommensurate
with $a$.

The simplest theoretical model for a CI transition is the 1D
Frenkel-Kontorova (FK) model \cite{Kontorova:39wq,ChLu01}. It
consists of a  chain of particles interconnected by springs,
representing an adsorbate layer, and placed in a periodic
potential describing the effects of a substrate. The potential
energy of a such system is given by

\begin{equation}
U=\sum_{n}
\left[\frac{\lambda}{2}\left(x_{n+1}-x_n-a\right)^2+V\left(x_n\right)\right],
\end{equation}
where $a$ is the equilibrium lattice spacing of the chain in the
absence of a potential and $\lambda$ is the stiffness. The
potential function $V\left(x_n\right)$ has periodicity $b$ and can
be approximated by a cosine function

\begin{equation}
V\left(x_n\right)=V_0 \left(1-\cos\left(\frac{2\pi
x_n}{b}\right)\right).
\end{equation}

When $V_0$ is sufficiently small, the adsorbate lattice will be
independent of the potential. This structure is called a
"floating" phase and the lattice spacing $\tilde{a} = \lim_{x \to
\infty}{(x_n-x_0)}/{n}$ of the adsorbate lattice can be an
arbitrary multiple of the substrate periodicity. In general, the
floating phase is incommensurate for almost all values of the
ratio $\tilde{a}/b$.

In the opposite limit, when the potential is very strong,
one can expect the lattice to relax into a commensurate structure
where the average lattice spacing of the adsorbed atoms is a simple
rational fraction of the period $b$. In the I phase, close to the CI
transition, it is energetically more favorable for the system to
form C regions separated by domain walls in which the springs are
stretched or compressed and the commensurate registry with the
potential is lost. These domain walls are usually called
discommensurations and the corresponding region in the phase diagram
within the I phase will be referred to as a modulated (M) phase. A
positive (negative) discommensuration leads to a reduction (increase)
in the density of adsorbate atoms and these regions are referred to
as light (heavy) walls.

The CI transition can be determined by examining the behavior of
the winding number $\tilde{ \Omega}={\tilde {a}/{b}}$ as a
function of $\Omega={a}/{b}$ for fixed $V_0$. In the continuum
version of the FK model the CI transition is second order with a
correlation length, identified as the domain wall separation,
which diverges logarithmical near the transition. If $V_0$ is
larger then a certain critical value, $V_c$, the system will be
commensurate for all the values of $\Omega$. The function
$\tilde{\Omega} \left(\Omega\right)$ has a staircase type of
appearance and thus for $V_0 < V_c$ it is called an incomplete
devil's staircase and for $V_0>V_c$ a complete devil's staircase
\cite{Bak:82be,Jensen:83ca}. If there are discontinuities or first
order jumps between commensurate states it is called a harmless
staircase~\cite{Villain:80ij}.

The FK model can be extended to two dimensions to describe, for
example, adsorbed layers on crystal surfaces. In its simplest
version the 2D FK model \cite{Bak:82be} describes the adsorbate
interactions by a pure harmonic potential in a periodic pinning
potential. Although the FK model takes into account topological
defects in the form of domain walls it leaves out plastic
deformations of the layer due to topological defects such as
dislocations. These defects are particularly important when the CI
transition occurs between two different crystal structures or in
presence of temperature fluctuations or quenched disorder.

They are automatically included in a full microscopic model
involving  interacting atoms in the presence of a substrate
potential. However, the full complexities of the microscopic model
limits the actual numerical computation to systems of relative
small sizes. The size effect could be very strong for CI
transition involving extended domains or topological defects.
Recently a phase field crystal model was introduced
\cite{Elder:04rq,Elder:02kz} that allows for both elastic and
plastic deformations in the solid phase. In this formulation a
free energy functional is introduced which depends on a density
averaged over microscopic times scales, $\Phi(\vec{r},t)$. The
free energy is minimized when $\Phi$ is spatially periodic
(\emph{i.e.}, crystalline) in the solid phase and constant in the
liquid phase. By incorporating phenomena on atomic length scales
the approach naturally includes elastic and plastic deformations,
multiple crystal orientations and anisotropic structures in a
manner similar to other microscopic approaches such as molecular
dynamics. However, the PFC method describes the density on a
diffusive and not the real microscopic times scales. It is
therefore computationally much more efficient. Thus this model
should provide a suitable description of the CI transition when
topological defects, domain walls and dislocations are present.

In this work we extend the 2D phase field crystal (PFC) model
\cite{Elder:04rq,Elder:02kz} to include the influence of an
external pinning potential. Such a model should provide a suitable
continuum description of lattice systems such as weakly adsorbed
atomic overlayers or 2D vortex lattices with pinning. The pinning
potential is chosen such that it induces CI transitions between
ground states of different symmetries in the model. The outline of
the paper is as follows. We first define the model and carry out
an analytic mode expansion to determine the crystal structure and
the location of the CI transition lines. We analyze these
transitions as a function of the lattice mismatch and strength of
the potential. Following this we carry out a numerical
minimization of the full free energy functional which gives good
agreement with the theoretical predictions and provides details on
the nature of the topological defects near the transition. Finally
we discuss the nature of the CI transitions and the relation of
our results to atomistic simulations for adsorbed overlayers.
\cite{Patrykiejew:01lm}.

\section{The Phase Field Crystal Model}

In the phase field crystal model \cite{Elder:04rq,Elder:02kz}, the
free energy functional is written as
\begin{equation}\label{one}
\mathcal{F} =\int d\vec r\left(\frac{ a\triangle T}{2}\Phi^2+u
\frac{\Phi^4}{4}+\frac{\Phi}{2}G(\nabla^2)\Phi\right),
\end{equation}
where $G(\nabla^2)=\lambda \left(q_0^2+\nabla^2\right)^2$, and its
eigenvalues can be related  to the experimental structure factor
of \emph{e.g.} Ar \cite{Elder:04rq}. Eq. (\ref{one}) describes a
crystal with a lattice constant of $2\pi/q_0$, while the elastic
properties can be adjusted by $\lambda$, $u$ and $q_0$. For
numerical calculation it is convenient to rewrite the free energy
in dimensionless units \cite{Elder:04rq,Elder:02kz} as
\begin{equation}
\vec{x}=\vec{r} q_0,\psi=\Phi\sqrt{\frac{u}{\lambda
q_0^4}},r=\frac{a\triangle T}{\lambda q_0^4},\tau=\Gamma\lambda
q_0^4t.
\end{equation}
In these units the free energy becomes
\begin{equation}
F = \frac{\mathcal{F}}{\lambda^2q^{8-d}/u}=\int
d\vec{x}\left[\frac{\psi}{2}\omega\left(\nabla^2\right)\psi+\frac{\psi^4}{4}\right],
\label{dimF}
\end{equation}
where $\omega\left(\nabla^2\right)=r+\left(1+\nabla^2\right)^2$.
Since $\psi$ is a conserved field it satisfies the following
equation of motion,
\begin{equation}
%\label{moteq}
\frac{\partial\psi}{\partial\tau}= \nabla^2 \frac{\delta F}{\delta
\psi} + \zeta = \nabla^2\left(\omega\left(\nabla^2\right)\psi
+\psi^3\right)+\zeta, \label{dimM}
\end{equation}
where $\zeta$ has zero mean,
 $\langle\zeta\left(\vec{x}_1,\tau_1\right)\zeta\left(\vec{x}_2,\tau_2\right)\rangle
=\mathcal{D}\nabla^2\delta\left(\vec{x}_1-\vec{x}_2\right)
\delta\left(\tau_1-\tau_2\right)$ and $\mathcal{D}$ is a constant.
Equations ~(\ref{dimF}) and (\ref{dimM}) have been used to study a
variety of phenomena involving elastic and plastic deformation
including grain boundary energies between misaligned grains,
buckling and dislocation nucleation in liquid phase epitaxial
growth, the reverse Hall-Petch effect in nano-crystalline
materials, grain growth and ductile fracture \cite{Elder:04rq}.

In 2D the free energy in Eq.~(\ref{dimF}) in the absence of
external potential is minimized by three distinct solutions for
the dimensionless field $\psi$; constant, stripes and dots. For
the purposes of this work only the latter solution is relevant.
This solution consists of a triangular distribution of density
maxima corresponding to a crystalline phase. In general this
solution can be written as
\begin{equation}
\psi(\bar r)=\sum_{n,m} a_{n,m}e^{i\vec G_{nm} \cdot\vec
r}+\bar\psi,
\end{equation}
where $\vec G_{nm}\equiv n\vec b_1 + m\vec b_2$, $\vec b_1$ and
$\vec b_2$ are the reciprocal lattice vectors and $\bar\psi$ is
the average value of $\psi$.  For a triangular lattice $\vec{b}_1$
and $\vec{b}_2$ can be written as
\begin{eqnarray}
\vec b_1 & = &\frac{2\pi}{a_t\sqrt{3}/2}\left(\sqrt{3}/2\hat
x+\hat y/2\right);\nonumber\\
\vec b_2 & = &\frac{2\pi}{a_t\sqrt{3}/2}\hat y,
\end{eqnarray}
where $a_t$ is the distance between nearest-neighbor (local)
maxima of $\psi$ (corresponding to the "atomic" positions). The
amplitudes $a_{m,n}$ and lattice spacing $a_t$ are determined by
minimizing the free energy functional.  For simplicity it is
useful to first consider a simple one-mode approximation (OMA) in
which only pairs $(n,m)$ that correspond to $|\vec{G}_{n,m}| =
2\pi/(a_t\sqrt{3}/2)$ are retained.  In this limit $\psi$ can be
written,

\begin{equation}
\psi_t=A_t[\cos(q_tx)\cos(q_ty/\sqrt{3})-
\cos(2q_ty/\sqrt{3})/2]+\overline{\psi},
\label{psiH}
\end{equation}
where $A_t$ is an unknown constant and $q_t=2\pi /a_t$.
Substituting Eq.~(\ref{psiH}) in Eq.~(\ref{dimF}) and minimizing with
respect to $A_t$ and $q_t$ gives,
$A_t=4(\overline{\psi}+(-15r-36\overline{\psi}^2)^{1/2}/3)/5$,
$q_t=\sqrt{3}/2$ and

\begin{equation}
\label{ft}
F^t_{min}/S=-\frac{1}{10}\left(r^2+\frac{13}{50}
\overline{\psi}^4\right)+\frac{\overline{\psi}^2}{2}
\left(1+\frac{7}{25}r\right)+\frac{4\overline{\psi}}{25}
\sqrt{-15r-36\overline{\psi}^2}
\left(\frac{4\overline{\psi}^2}{5}+\frac{5}{3}\right).
\end{equation}

As shown in Ref.~\onlinecite{Elder:04rq} this approximation is
valid in the limit of small $r$.  In the next section an external
pinning potential with square symmetry and incommensurate with
will be introduced in the PFC model. As will be shown , depending
on the strength of the external potential and the lattice
mismatch, a commensurate incommensurate transition would occur.

\section{Phase Diagram With External  Potential}

In the PFC model ,  an external potential $V$ is introduced by
adding a term coupling $V$ linear to  $\psi$ in the free energy
functional given in Eq.~(\ref{dimF}), \emph{i.e.},

\begin{equation}
F = \int
d\vec{x}\left[\frac{\psi}{2}\omega\left(\nabla^2\right)\psi+\frac{\psi^4}{4}+\psi\,V\right].
\label{pinF}
\end{equation}

In this study, we consider an external potential of square
symmetry which is distinct from the symmetry of the triangular
lattice in the absence of the external potential (\emph{e.g.}, Eq.
\ref{psiH}).

\begin{equation}
V=V_0\cos(q_{s}x)\cos(q_{s}y) \label{pin1}.
\end{equation}
where $q_{s}=2\pi/(a_s\sqrt{2})$. We define the relative mismatch
$\delta_\mathrm{m}$ between the external potential and adsorbed
monolayer as

\begin{equation}
\delta_\mathrm{m}=1-2\pi/a_s %\delta_\mathrm{m}=1-2\pi/a_s
\end{equation}
With the above external potential, the PFC model could describe
for example an adsorbed layer on the (100) face of an fcc crystal
\cite{Patrykiejew:01lm}.

To understand the influence of $V$ on the minimum energy solution,
we can again Fourier analyze the equilibrium density. Taking into
account both the intrinsic triangular symmetry and the external
potential of square symmetry, the system is best described by a
combination of hexagonal and square modes. In this case the
corresponding hexagonal-square mode approximation (HSMA) can be
written as

\begin{eqnarray}
\label{psihs}
\psi_{hs}&=&A_{1}\cos(q_{s}x)\cos(q_{s}y)+A_{2}
\left(\cos(2q_{s}x)+\cos(2q_{s}y)\right)\nonumber\\
&+&A_t[\cos(q_tx)\cos(q_ty/\sqrt{3})-\cos(2q_ty/\sqrt{3})/2]+
\overline{\psi}.
\end{eqnarray}

This Fourier expansion includes the basic mode for the triangular
lattice as done in  Eq. \ref{psiH} and the first two harmonics for
the commensurate modes with square symmetry. In general, they
describe an incommensurate phase corresponding to an triangular
phase distorted by the square symmetry external potential.
However, when the mismatch is sufficiently large  and/or the
strength of the external potential is sufficiently strong, the
solution that minimize the free energy will correspond to a
vanishing value for$A_t$ and a commensurate phase.

The free-energy density in the HSMA approximation is given by

\begin{eqnarray}
\label{hsma} F^{hs}/S&=&\frac{9}{256} A_{1}^4 +\frac{9}{16}
A_{2}^4  + A_{1}^2\left(\frac{1}{8}
+\frac{r}{8}+\frac{3\overline{\psi}^2}{8}
-\frac{1}{2}q_{s}^2+\frac{1}{2}q_{s}^4\right)+A_{2}^2\left(\frac{1}{2}+
\frac{3}{2} \overline\psi^2
+r\frac{1}{2}+8q_s^4-4q_s^2\right)\nonumber\\&+&\frac{9}{16}A_t^2\left(\frac{1}{4}A_1^2+A_2^2\right)+
A_{1}^2\left(\frac{9}{16}
A_{2}^2+\frac{3}{4}A_2\overline\psi\right)
+\frac{V_0A_{1}}{4}+F^{t}/S.
\end{eqnarray}

The coefficients $A_t$, $A_{1}$ and $A_{2}$ are unknown and must be chosen to
minimize the free-energy density.

\begin{figure}[!h]
\begin{center}
\subfigure[]
{\includegraphics[width=70mm,clip=true]{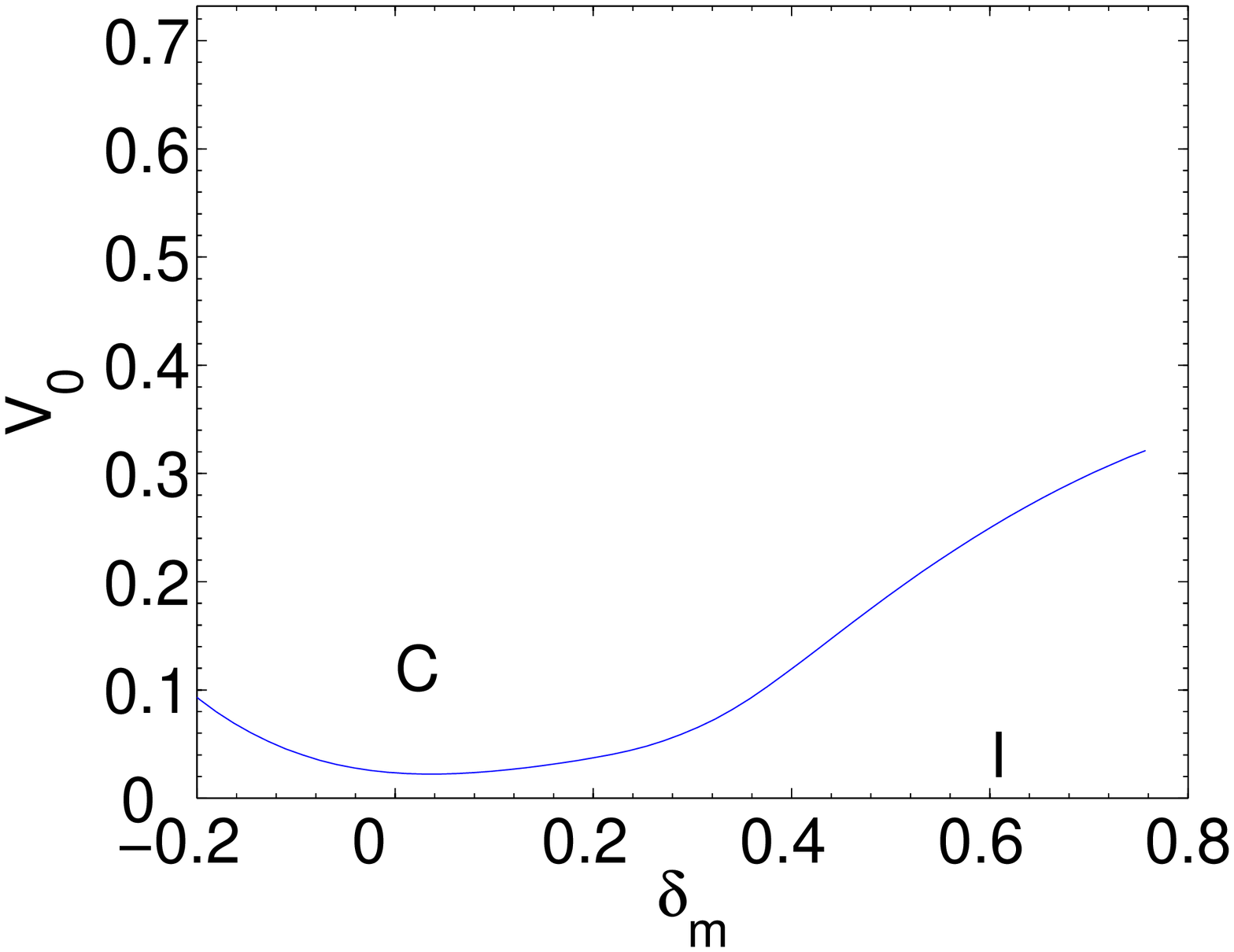}\label{anpd}}
\subfigure[]
{\includegraphics[width=70mm,clip=true]{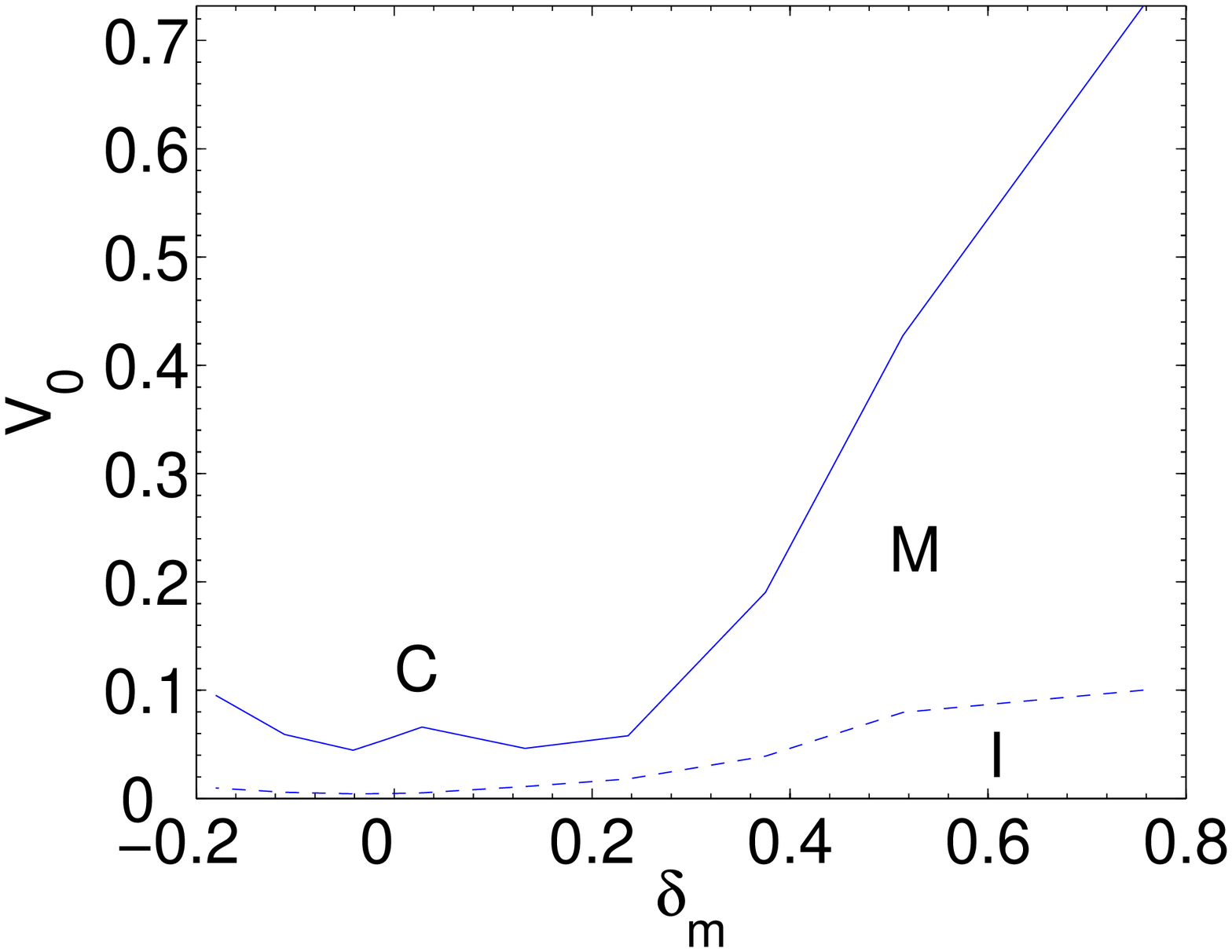}\label{nupd}}
\end{center}
\caption{\label{pd} (a) The phase diagram calculated analytically
using the HSMA approximation for the free energy. The region denoted
by C is the commensurate phase and I denotes the incommensurate
phase. (b) The corresponding phase diagram determined numerically as
explained in Sec. \ref{numerics}. The dashed line corresponds to a
crossover regime between the fully incommensurate I and the
modulated M phases. See text for details.}
\end{figure}

The analytic expressions for the free energies can now be used to
obtain the phase diagram of the pinned PFC model as a function of
the pinning strength $V_0$ and the mismatch $\delta_{\rm m}$. To
this end Eq. (\ref{hsma}) can be used to determine the critical
value of $V_0$,$V_c$, at which the CI transition occurs,
identified by the point where the amplitude of the triangular
phase vanishes, \emph{i.e.}, when $A_t(V_0,\delta_{\rm m})=0$. The
results are shown in Fig. \ref{pd}(a). Within the HSMA the CI
transitions is discontinuous and becomes continuous only for
infinite $\delta_{\rm m}$. In the next section we will compare the
analytically obtained phase diagram in the HSMA approximation with
a full numerical minimization of the total free energy.

\section{Numerical Results}
\label{numerics}

\subsection{Energy Minimization}
While the HSMA yields a good qualitative understanding for the phase
diagram describing the CI transition, it is not quantitatively
accrate, particularly near the transition, since the HSMA cannot
describe the structure of the domain walls and other possible
topological defects. In this section, we desribe the full numerical
investigation of the CI transition in the PFC model. This is
obtained by direct minimization of the free energy functional
without any assumed form for the density field.

In the presence of the external potential, the equation of motion
for $\psi$ becomes,

\begin{equation}
\label{pineq} \frac{\partial\psi}{\partial\tau}
=\nabla^2\left(\omega\left(\nabla^2\right)\psi+\psi^3 +V
\right)+\zeta .
\end{equation}

The minimal energy numerical solutions for the equilibrium states
of $\psi$ were obtained by solving $\delta F/\delta \psi=0$ using
a simple relaxational method similar to the usual molecular
dynamics annealing scheme. The noise term $\zeta$ in
Eq.~(\ref{pineq}) is only used as an annealing  tool to escape
from any metastable states. It is applied for a limited period of
time only and then set back to zero. For the remainder of the
time, Eq.~(\ref{pineq}) was solved using a simple Euler algorithm,
\emph{i.e.},
\begin{equation}
\label{deqm} \psi_{n+1,i,j}=\psi_{n,i,j}+\Delta \tau
\nabla^2\left(\left[\left(r+\left(1+\nabla^2\right)^2\right)\psi_{n,i,j}
+\psi^3_{n,i,j}+V_{n,i,j}\right]\right),
\end{equation}
where the Laplacian operator $\nabla^2$ represents the lattice
second order derivatives. A so-called spherical Laplacian
approximation \cite{Elder:04rq} was used for $\nabla^2$. For a
thorough discussion on solving solving differential equations
numerically using stencils, see Ref.~\cite{Patra:05ax}.

Eq.~(\ref{deqm}) was solved on a $512\times 512$ grid with the
spatial discretization $dx=1$ and time step $\Delta\tau=0.02$
using periodic boundary conditions. The parameters $r$ and
$\bar\psi$ were chosen to correspond to a crystalline region of
phase space, \emph{i.e.}, $r=-1/4$ and $\bar\psi=-1/4$. However, a
hexagonal lattice cannot be fitted in a square geometry and in
order to satisfy the periodic boundary conditions, the lattice
will be distorted and it may exhibit domains of different
orientations separated by walls, thereby giving a free-energy
density higher than that of the ground state. The value of $dx$
also influences the free-energy density when lattice constant is
small. We have checked the finite-size effects for our lattice in
the absence of pinning potential, the relative difference between
free-energy density $F^n/S$ and the OMA is only about $0.56\%$.

\subsection{Phase Diagram and Ground State Configurations}

\begin{figure}[!h]
\begin{center}
\subfigure[$V_0=0.00$]{\includegraphics[width=43mm,clip=true]{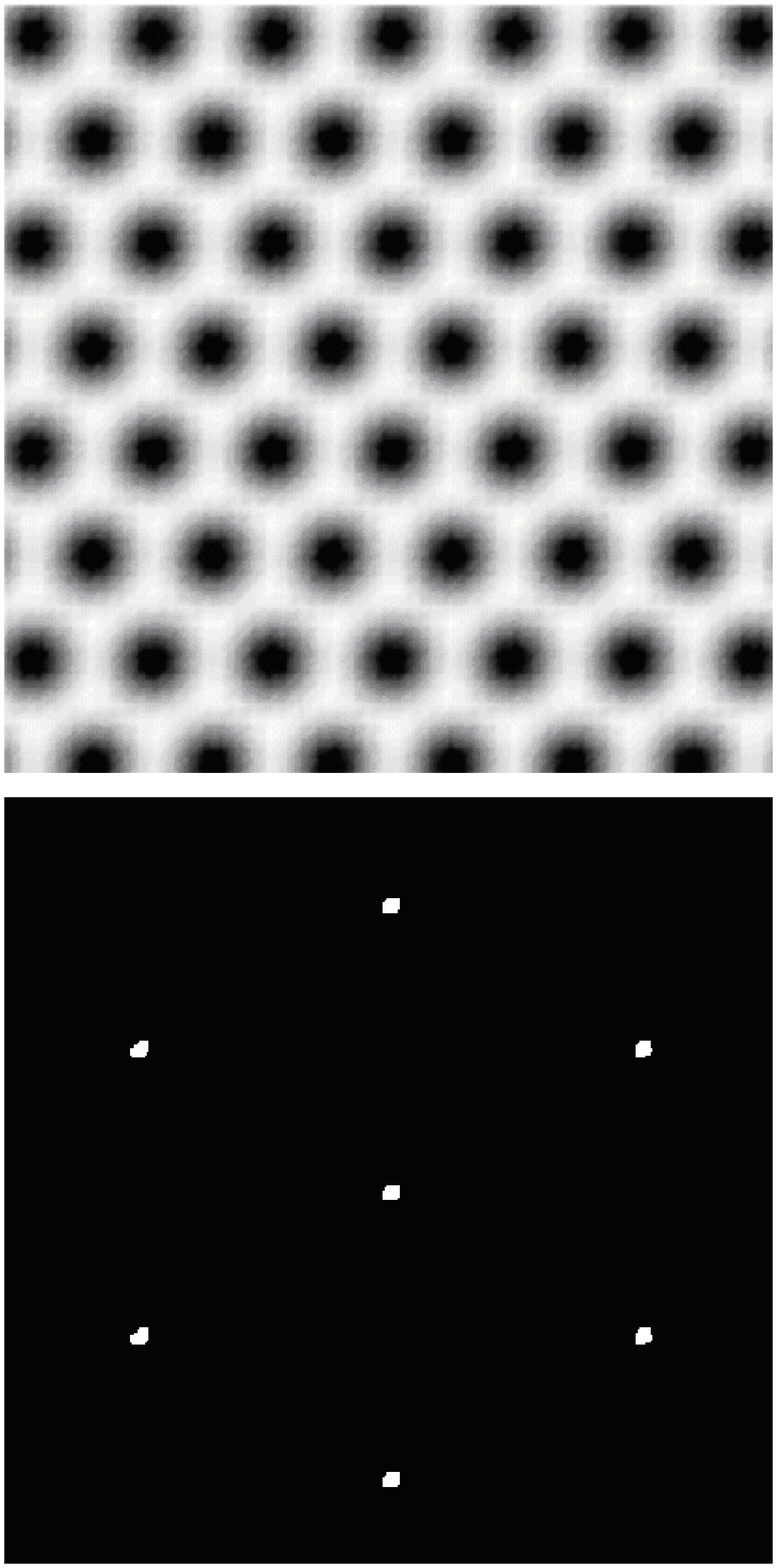}\label{v=0.00}}
\subfigure[$V_0=0.02$]{\includegraphics[width=43mm,clip=true]{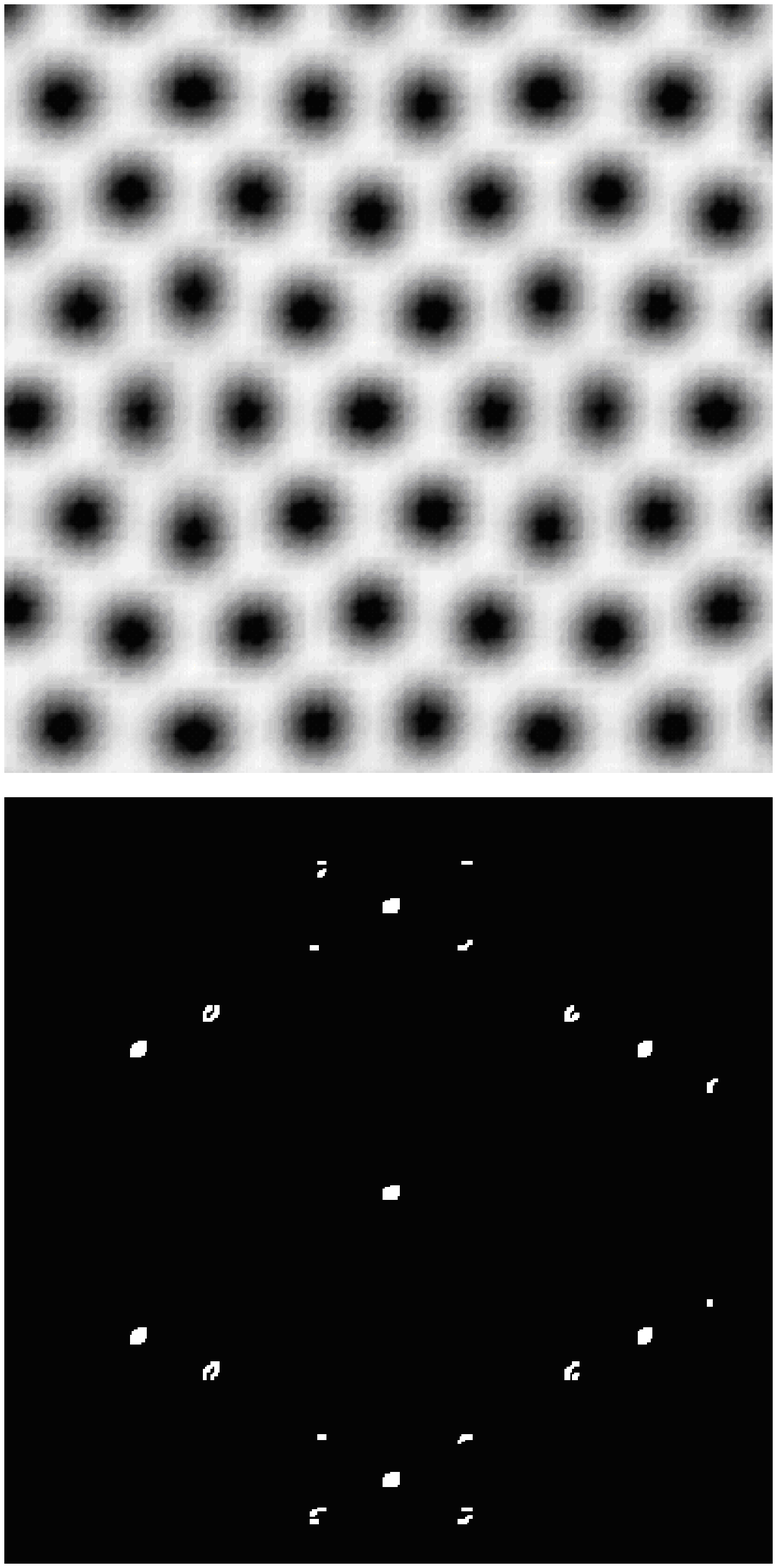}\label{v=0.02}}
\subfigure[$V_0=0.04$]{\includegraphics[width=43mm,clip=true]{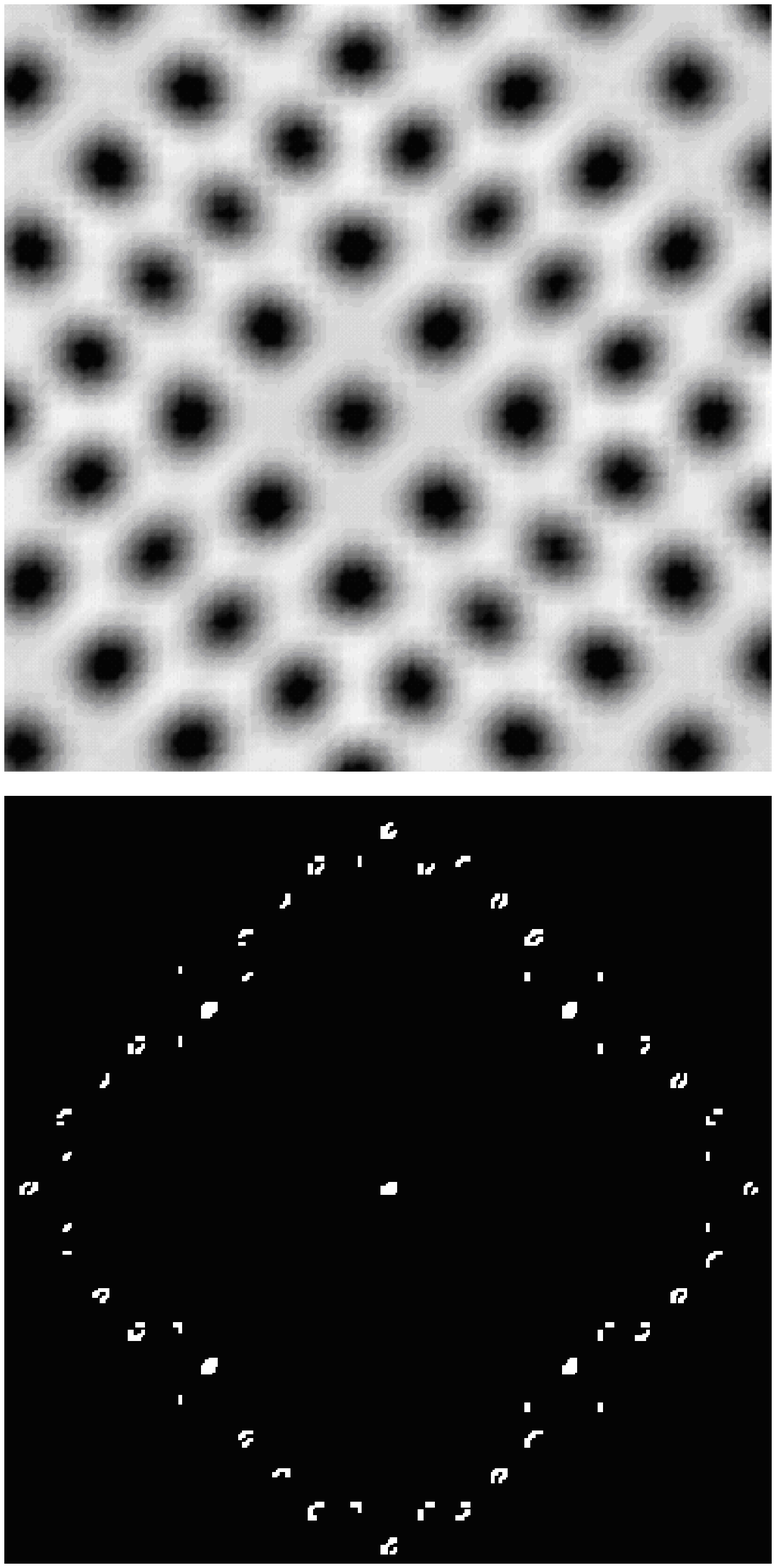}\label{v=0.04}}
\subfigure[$V_0=0.05$]{\includegraphics[width=43mm,clip=true]{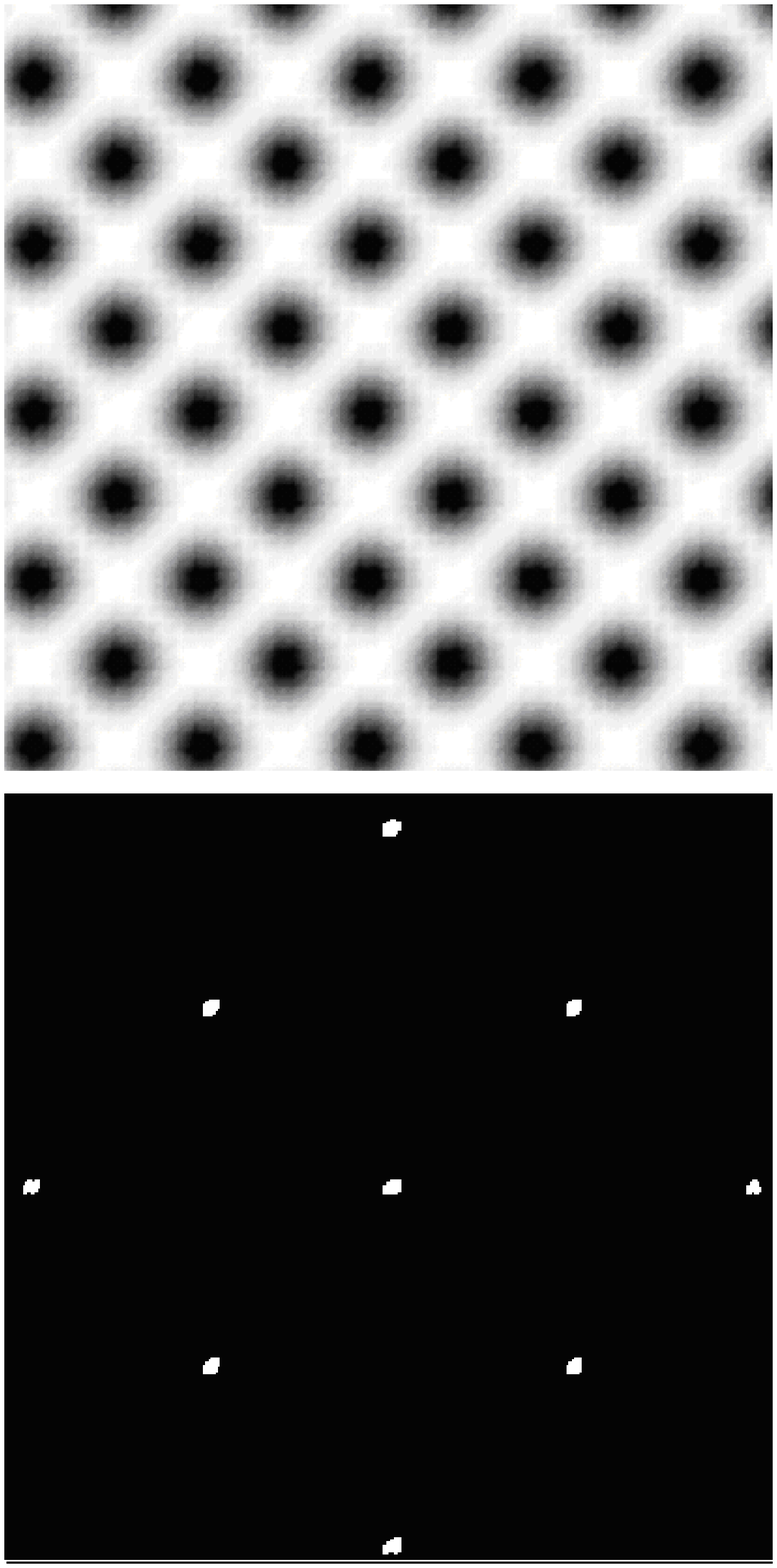}\label{v=0.05}}
\end{center}
\caption{\label{evolution}Snapshots of ground states and the
corresponding structure factors (lower figures) showing the
transition of the system from a triangular lattice for zero
pinning to a square lattice for $\delta_\mathrm{m}=0.14$ (only a
small part of the lattice is shown here).}
\end{figure}

The free energy density and the structure factor $S(|\vec
k|)=S(k)=|\tilde \psi (|\vec k|)|^2 $ were calculated numerically
for several different values of $2\pi /a_s$ as allowed by the
periodic boundary conditions. In Fig. \ref{evolution} a set of
configurations for the model with increasing amplitude of the
pinning potential $V_0$ are shown. As expected for $V_p=0$ the
ground state has perfect triangular symmetry. With increasing
amplitude of the square pinning potential the configurations
become spatially distorted and eventually the system undergoes a
transition to a square lattice.

\begin{figure}[!h]
\begin{center}
\subfigure[$V_0=0.04440$]{\includegraphics[width=50mm,clip=true]{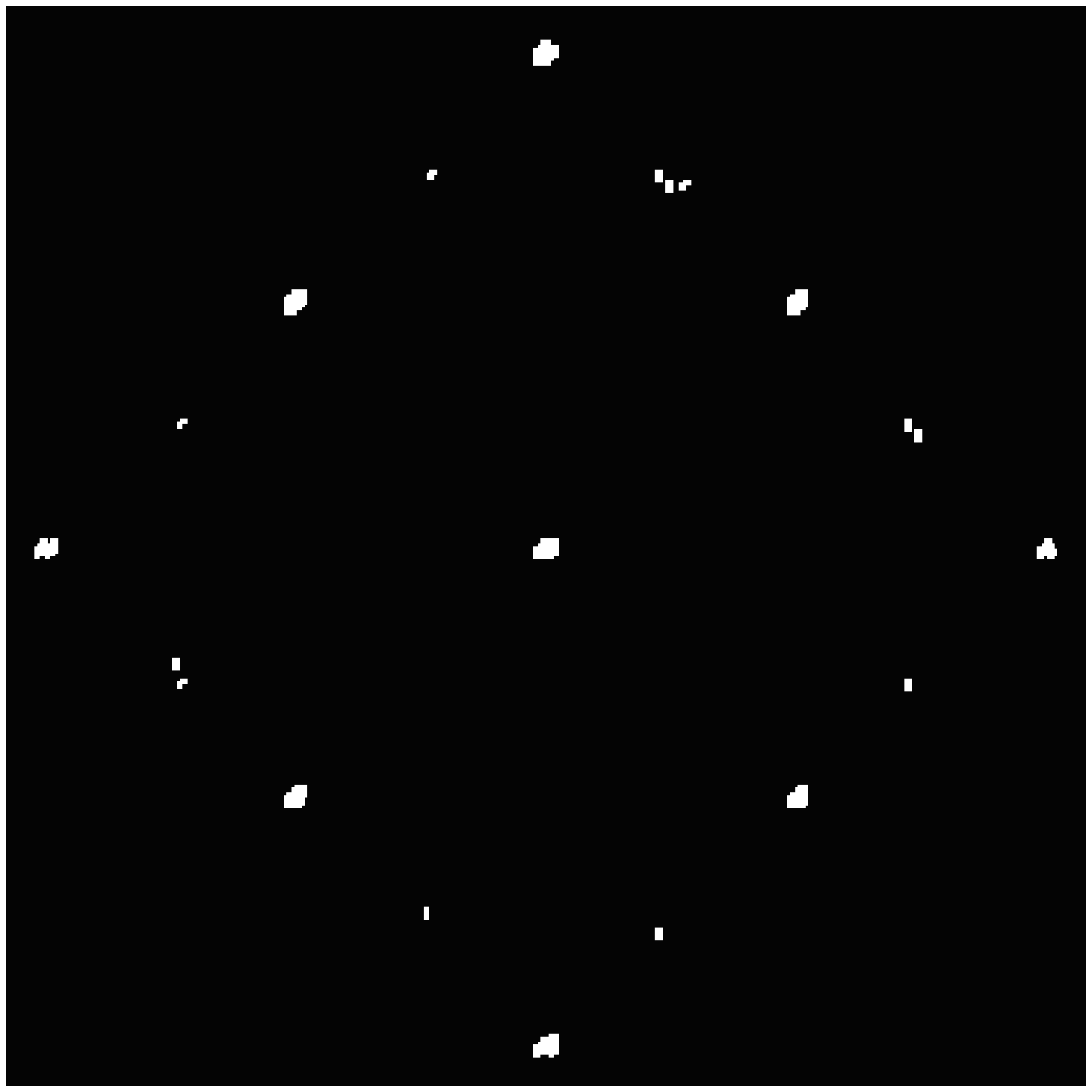}\label{v=0.001}}
\subfigure[$V_0=0.04455$]{\includegraphics[width=50mm,clip=true]{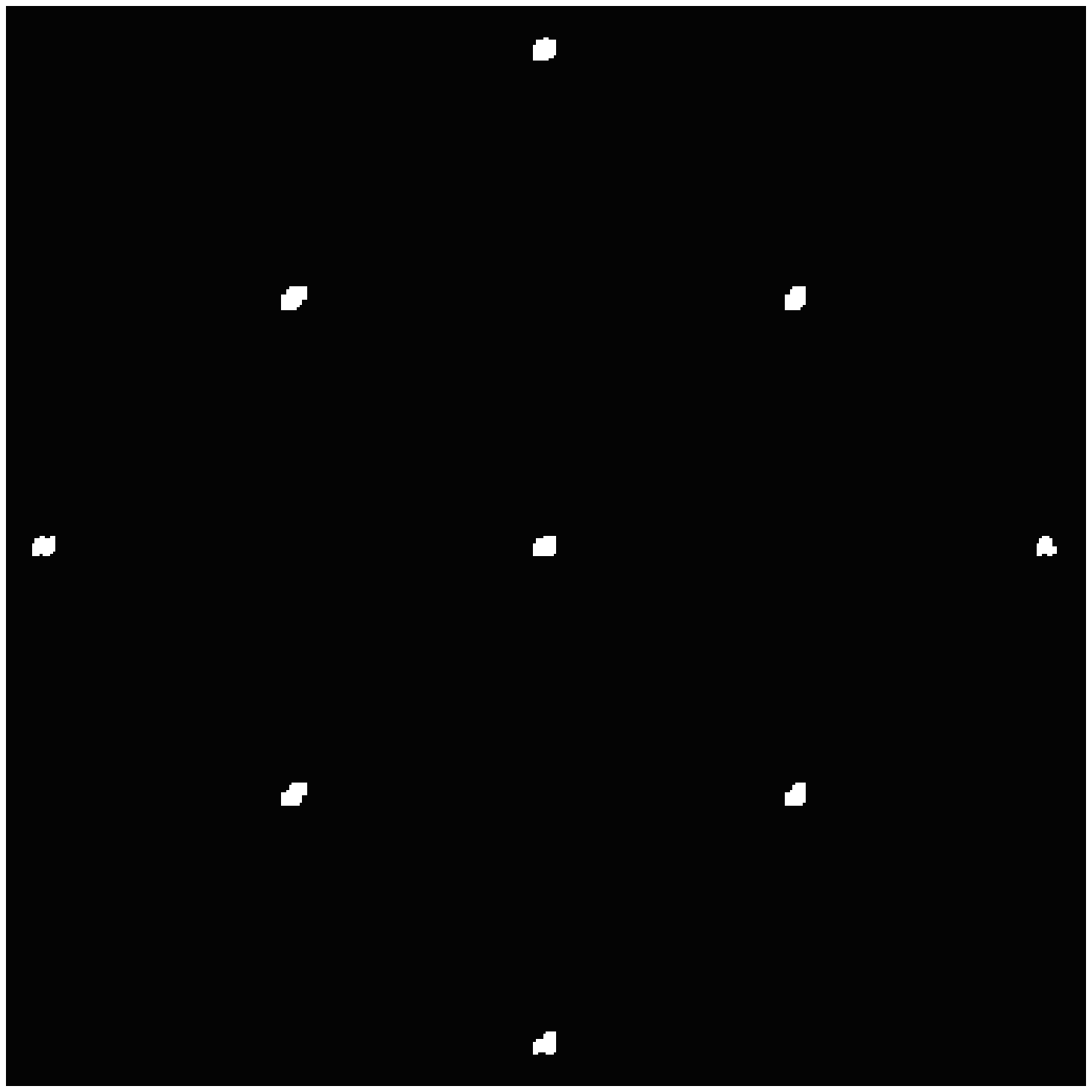}\label{v=0.002}}
\end{center}
\caption{\label{strev05}Splitting of the peaks of the structure
factor with decreasing pinning potential in the vicinity of the CI
transition for $\delta_\mathrm{m}=0.14$. The splitting is
discontinuous, suggesting a first-order like transition.}
\end{figure}

\begin{figure}[!h]
\begin{center}
\subfigure[$V_0=0.4275$]{\includegraphics[width=50mm,clip=true]{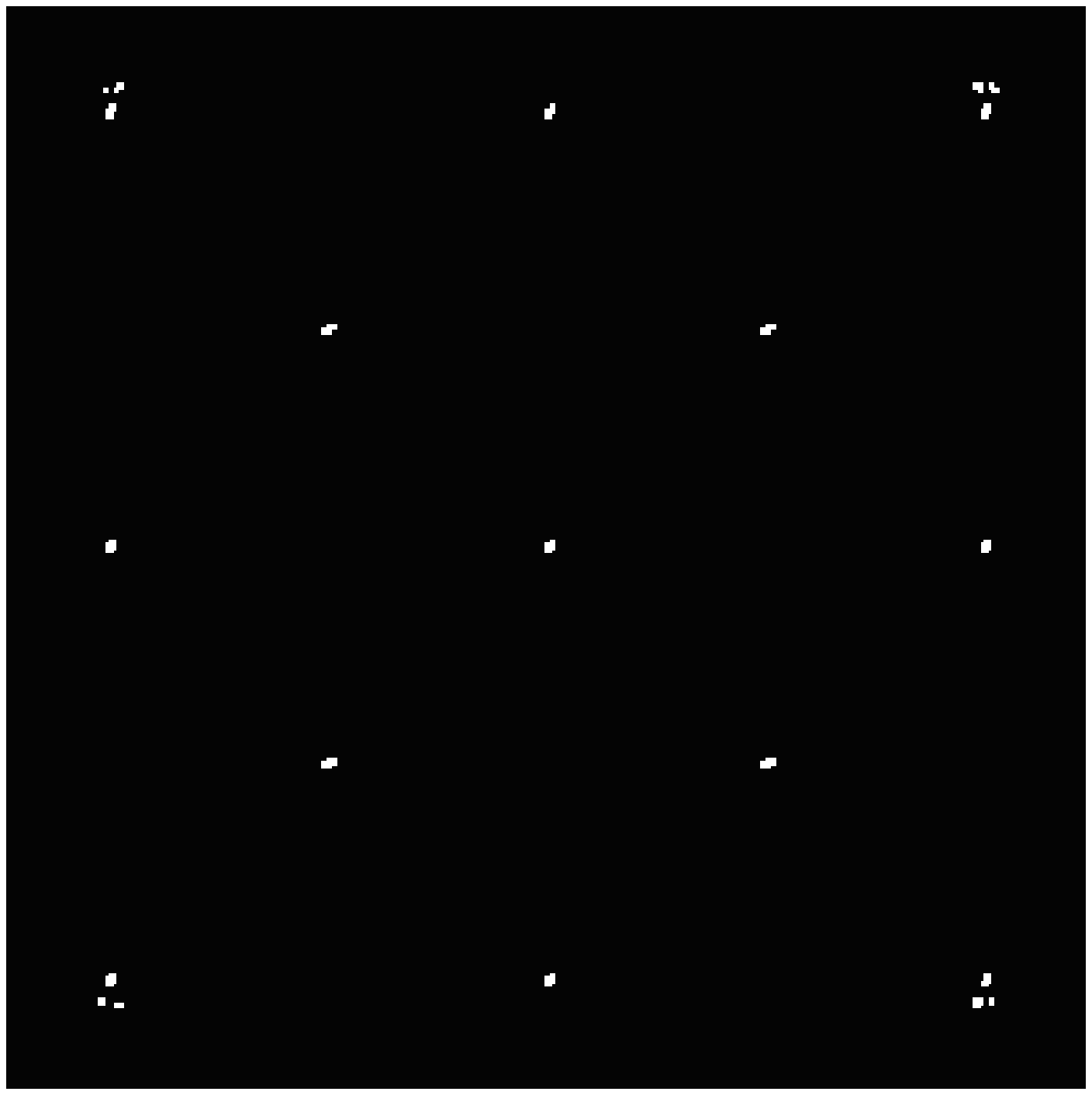}\label{v=0.001}}
\subfigure[$V_0=0.4455$]{\includegraphics[width=50mm,clip=true]{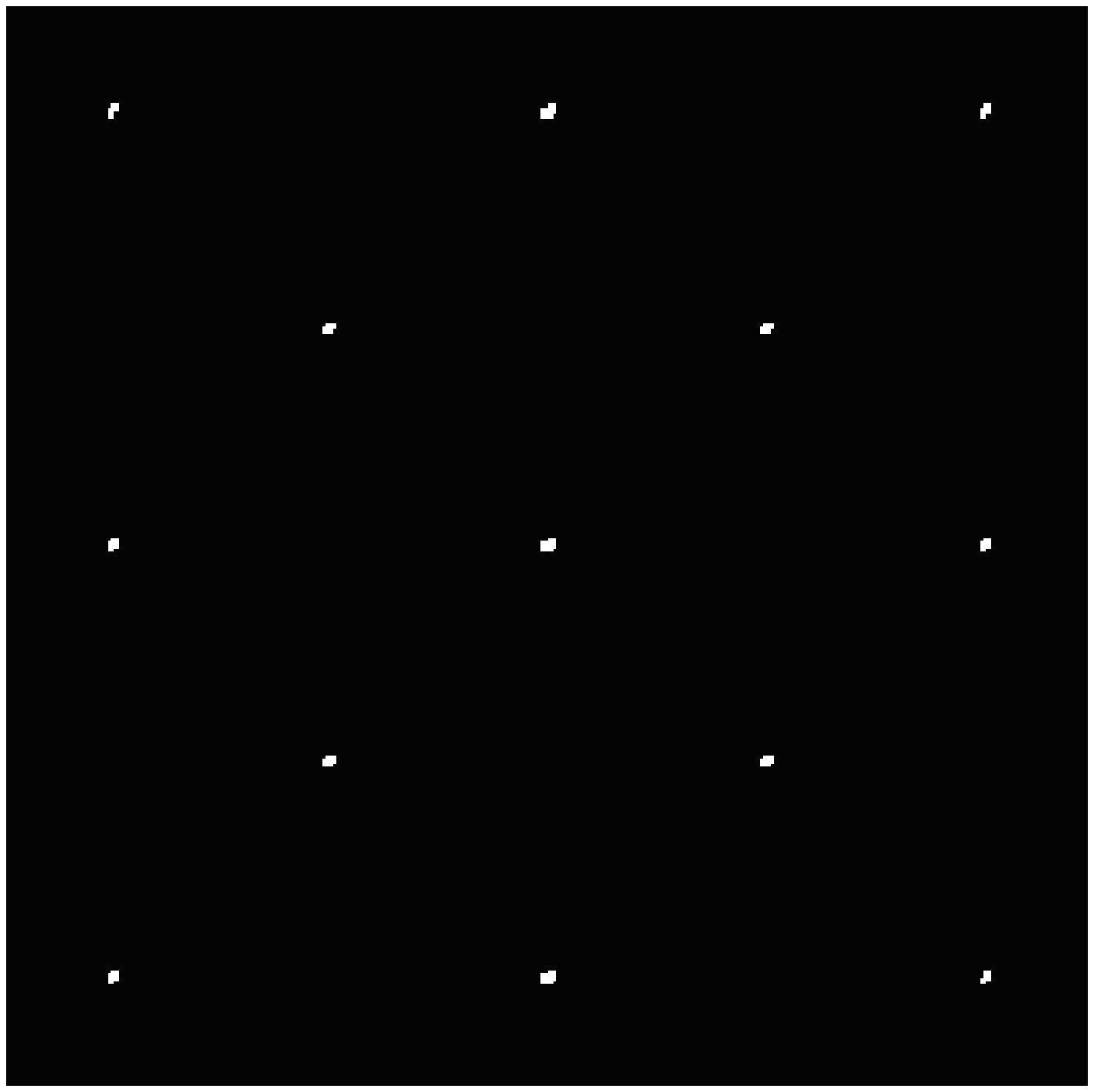}\label{v=0.002}}
\end{center}
\caption{\label{strev02}Splitting of the peaks of the structure
factor with decreasing pinning potential in the vicinity of the CI
transition for $\delta_\mathrm{m}=0.52$. The splitting is much
smaller than in Fig. 3 but remains discontinuous. See text for
details. }
\end{figure}

We define the position of the CI transition by analyzing the
structure factor $S(k)$. The transition occurs when the peaks of
the structure factor that correspond to the square lattice begin to
split as shown in Figs \ref{strev05} and \ref{strev02}.

\begin{figure}[!h]
\begin{center}
\subfigure[$\delta_\mathrm{m}=0.52$]{\includegraphics[width=38mm,clip=true,angle=-90]{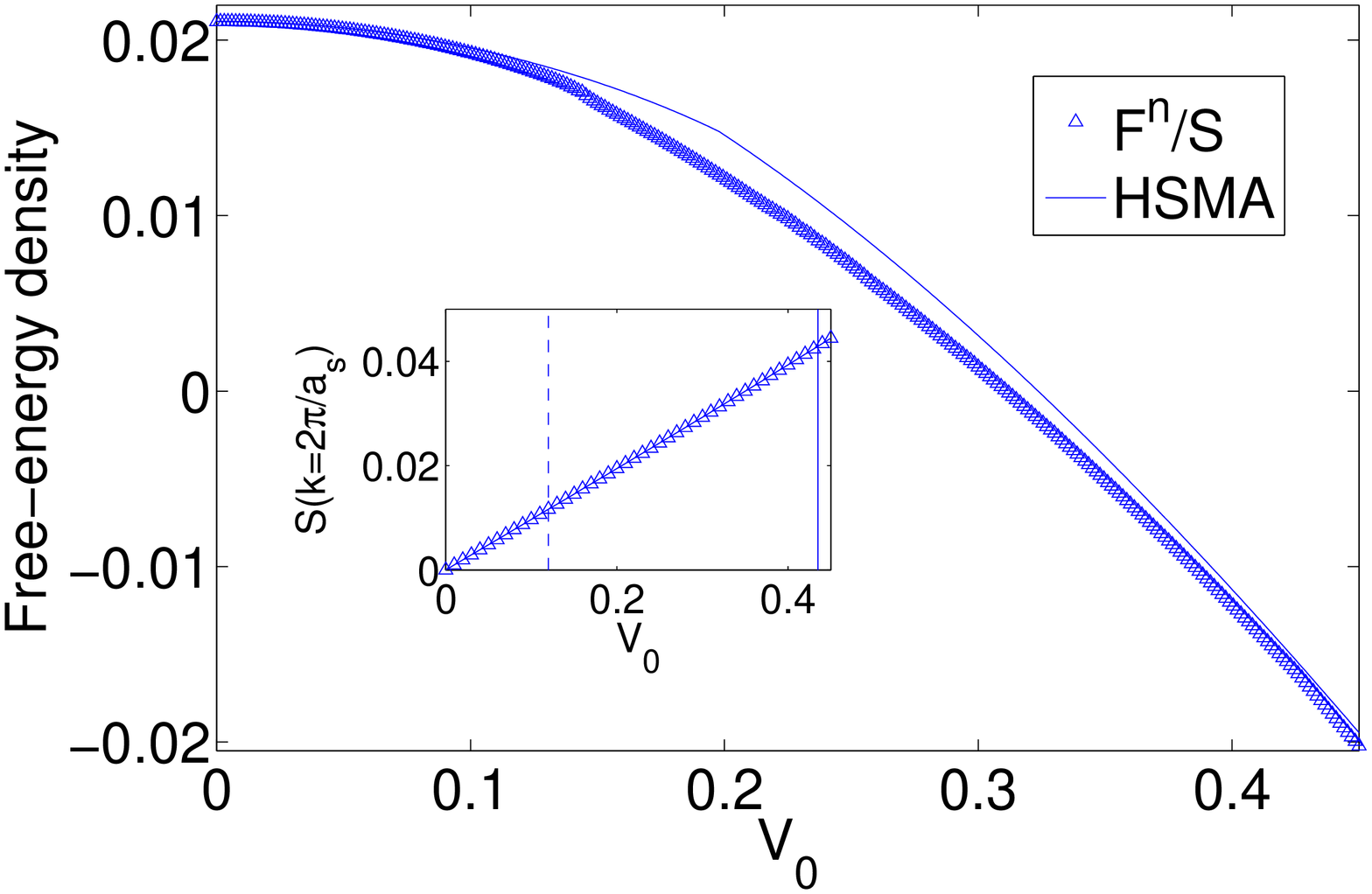}\label{en0.48}}
\subfigure[$\delta_\mathrm{m}=0.14$]{\includegraphics[width=38mm,clip=true,angle=-90]{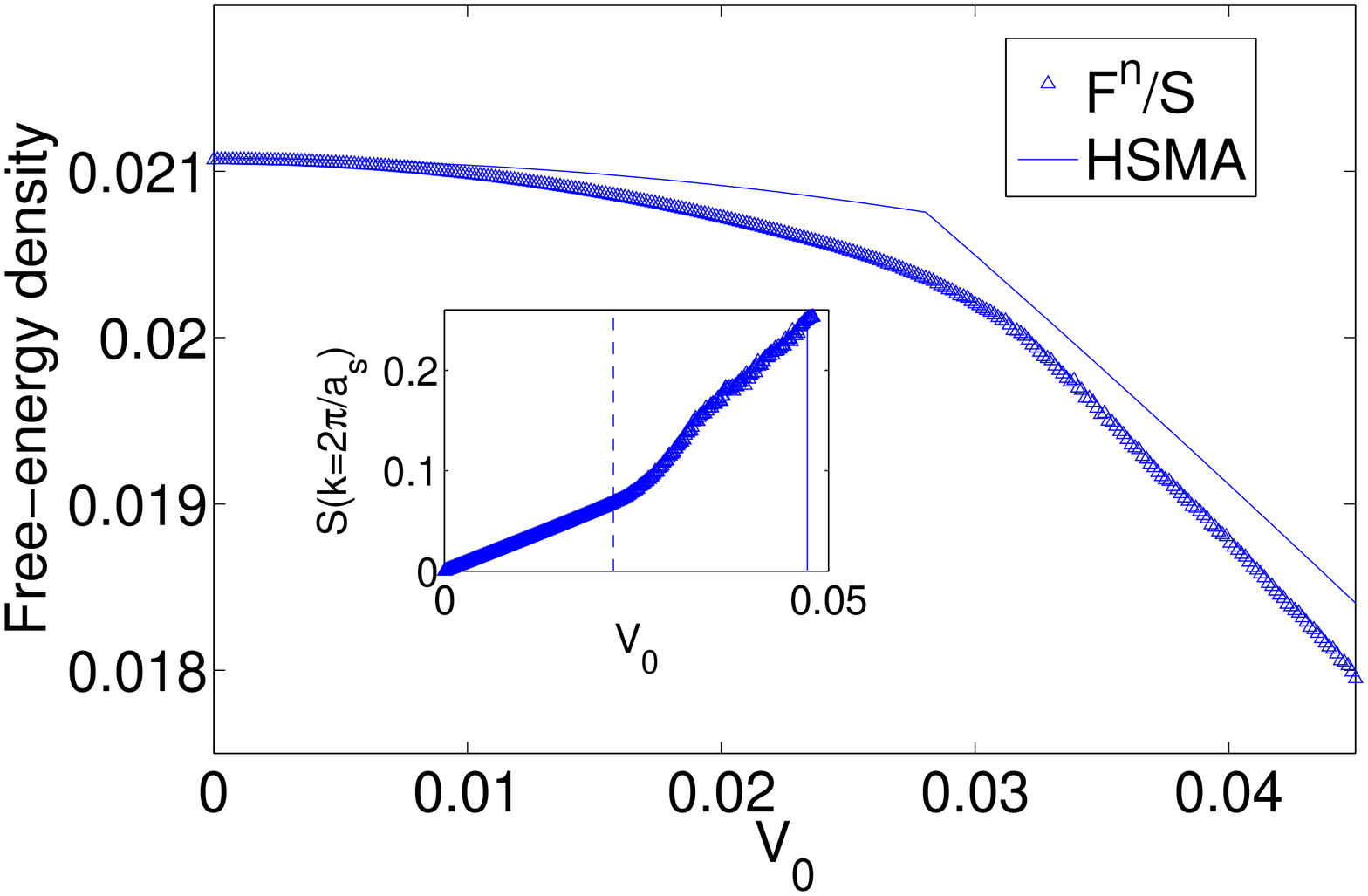}\label{en0.86}}
\subfigure[$\delta_\mathrm{m}=-0.01$]{\includegraphics[width=38mm,clip=true,angle=-90]{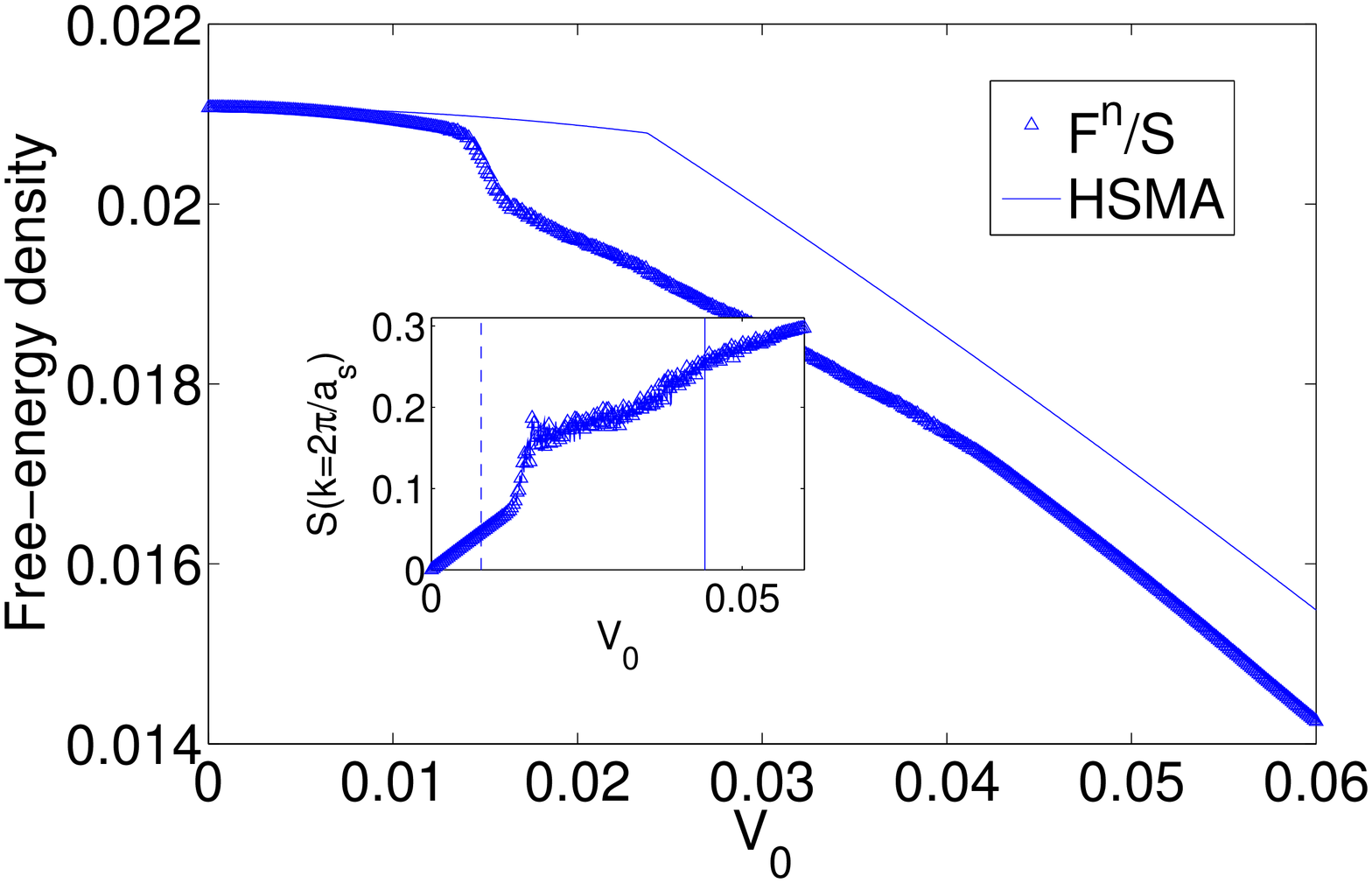}\label{en1.00}}
\end{center}
\caption{\label{enop}Comparison of the free-energy densities
between analytical (HSMA) and numerical results for different
values of $\delta_\mathrm{m}$. The inset represents intensity of
the peaks of the structure factor corresponding to the reciprocal
vectors of magnitude $k=2\pi/a_s$. The continuous vertical lines
represent the position of the CI transition, while the dashed
lines marks the cross-over regime in the I phase.}
\end{figure}

Next the approximate analytic solutions for the free energy are
compared to the numerical ones as a function of pinning potential
for fixed mismatch. As seen in Fig.~\ref{enop} the free-energy
density decreases as a function of the pinning strength. For small
values of $\delta_{\rm m}$ the transition to square symmetry
occurs at relatively weak pinning strengths and is first-order.
For larger values of $\delta_{\rm m}$ the free energy decreases
slowly and the CI transition appears continuous (see
Fig.~\ref{enop}(a)).  However, we have checked the finite-size
dependence up to systems of linear size 1024 and find that for up
to at least $\delta_\mathrm{m}=0.52$ the transition remains
first-order. We note that a first-order CI transition is in fact
consistent with theoretical predictions based on models of
interacting domain walls on a pinning potential with square
symmetry \cite{Rys01}. In Fig.~\ref{enop} we can also see that the
analytical HSMA gives reasonable agreement with the numerical
results, especially for large misfit where domain structures are
not as predominant.

Finally, we determined the onset of the modulated (M) phase within
the I phase by the value of $V_0$ where the peaks in the structure
factor corresponding to the triangular symmetry split in multiple
peaks. In the present model the M phase corresponds to regions
commensurate with the pinning potential, but separated by heavy
domains walls of excess of local density (see also Fig.
\ref{voro}). The dashed lines indicates the threshold value where
domain walls appear in the system, shown  in Figs. \ref{pd} and
\ref{enop}(a)-(c) (in the insets) by dashed lines.

\subsection{Voronoi Analysis of Domain Walls}

\begin{figure}
\begin{center}
\includegraphics[width=60mm,clip=true,angle=-90]{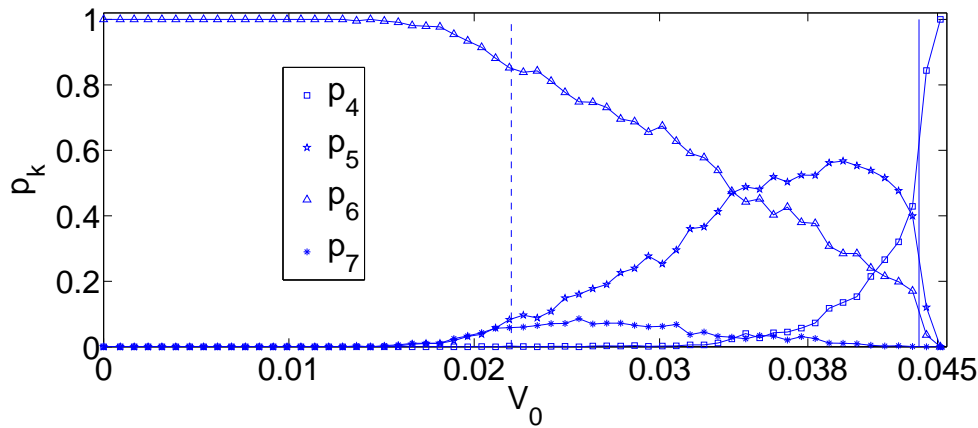}
\caption{\label{vorcon} Relative contributions of the Voronoi
polygons with $k$ NNs ($p_k$, $k=4,5,6,7$) as a function of the
pinning strength, with $\delta_{\rm m}=0.14$. The continuous line
represents the CI transition, while the  the dashed line marks
cross-over regime within the I phase.}
\end{center}
\end{figure}

\begin{figure}[!h]
\begin{center}
\subfigure[]{\includegraphics[width=50mm,clip=true]{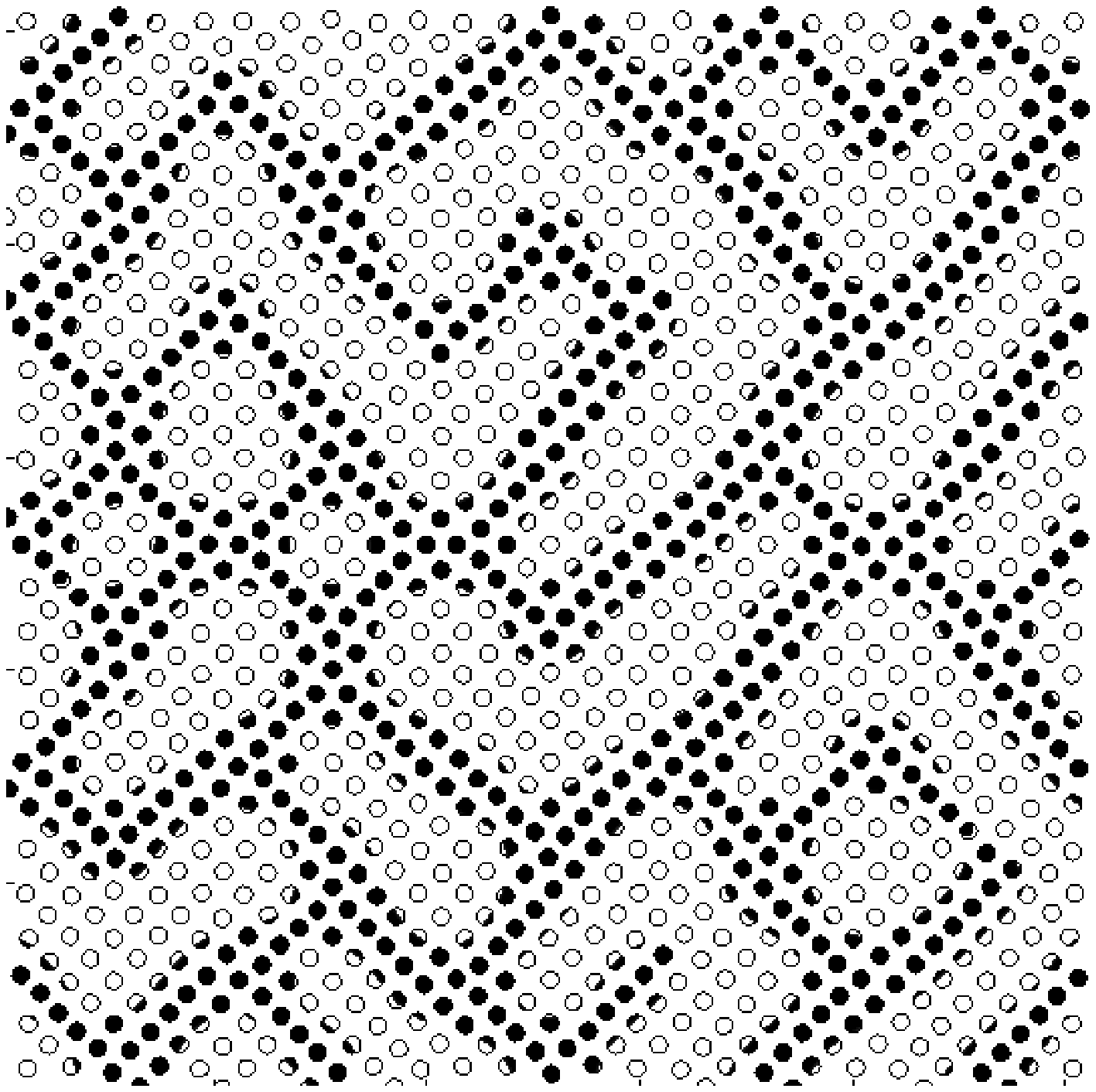}
\label{dom} } \subfigure[]
{\includegraphics[width=69.5mm,clip=true]{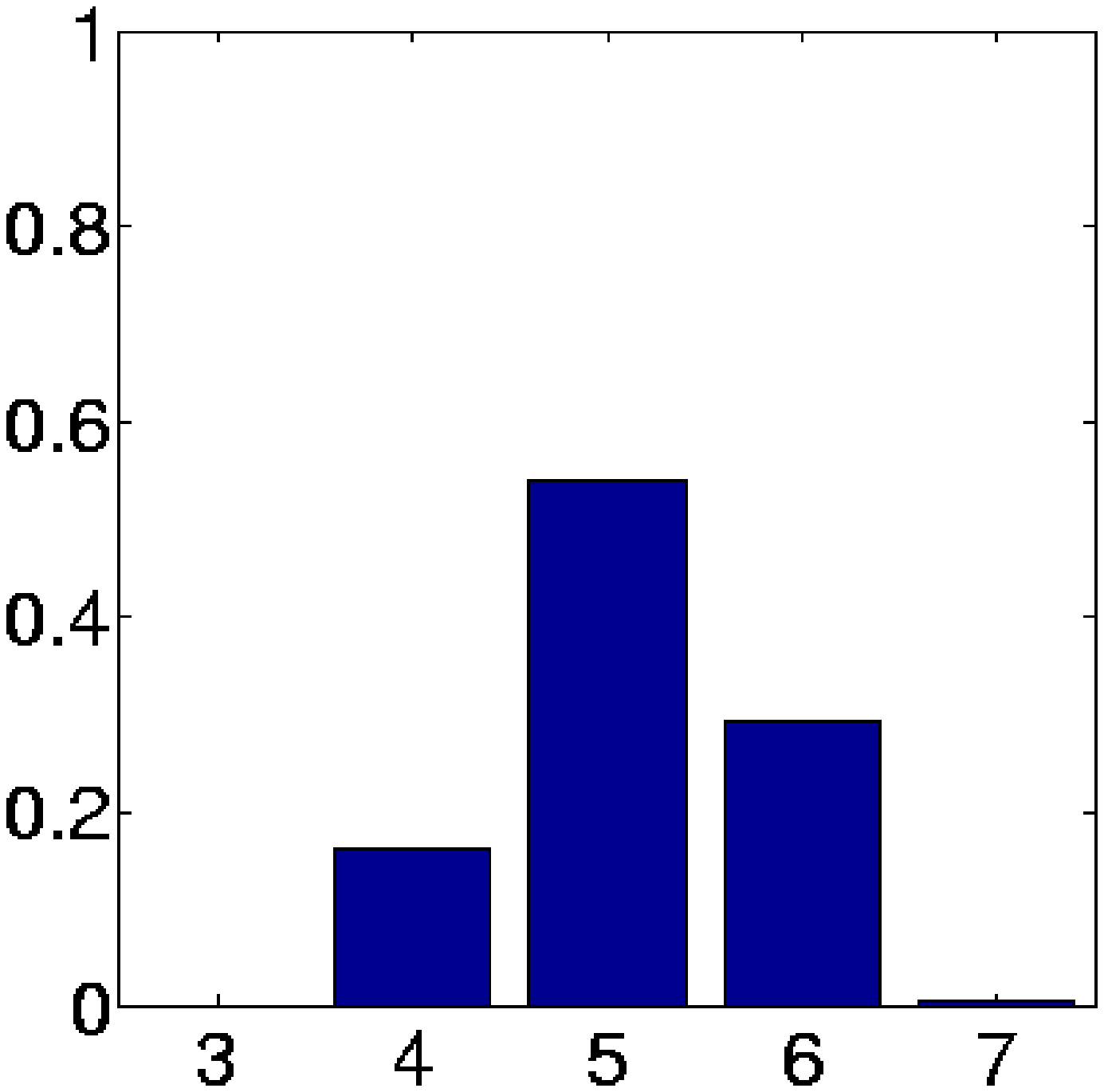} \label{htsg} }
\end{center}
\caption{\label{voro}(a) Example of modulated phase obtained from
system characterized by $\delta_\mathrm{m}=0.14,V_0=0.043$. The
dark areas represent the I regions, \emph{i.e.} heavy walls, while
the light areas are C regions. The positions of the 'particles'
are superimposed on the density maxima. (b) The contribution of
Voronoi polygons in this state. See text for details.}
\end{figure}

It is of further interest to analyze the nature of the spatial
configurations and domain walls in the I phase and in the
transition region. A convenient way to quantify the defects is to
use the Voronoi analysis \cite{Shinoda:98by,Flekkoy:00ne}. By
definition, the Voronoi cell of any particle contains all the
points that are closer to it  than any other particle, \emph{i.e.}
it defines the Wigner-Seitz cell for each particle. The resulting
cells are polygons with N sides that represents  the number of
nearest neighbors (NNs).  In the present case this is particularly
useful since very close to the transition region there is some
variation in the positions and sizes of the density maxima,
identified as effective "particles" in our phase field crystal
model. In Fig.~\ref{vorcon} we show the results of Voronoi
analysis of our data as a function the pinning strength for a
fixed mismatch $\delta_{\rm m}=0.14$. For weak pinning, Voronoi
cells with N=6 dominate as expected. Closer to the CI transition
line (shown with a vertical line) contributions from N=5 becomes
important and in the immediate vicinity of the CI transition
virtually all Voronoi cells have N=4.
The Voronoi cell analysis can be used to identify  the location
and nature of the domain walls. Close to the CI transition, the
system is composed of regions commensurate with the pinning
potential but separated by heavy walls. These configurations
appear spontaneously and have been observed also in Monte Carlo
simulations of overlayers of atoms interacting via the
Lennard-Jones potential adsorbed on the (100) face of an fcc
crystal \cite{Patrykiejew:01lm}. In Fig.~\ref{voro} we show
results for the present model in the corresponding region. The
density of maxima is smoothed with a gaussian function with width
of approximately $3a_s$.  As can be seen in Fig.~\ref{dom}, the
system indeed comprises commensurate regions separated by walls.
The regions of high density indicate "heavy" domain walls with
excess of "particles" compared to the commensurate regions . In
these regions the "particles" are plotted with black circles while
in the commensurate regions with white circles. In Fig.~\ref{htsg}
we show the contribution of different Voronoi polygons in this
state. The fraction of polygons  with N=5 ($p_5$) is $0.53$, while
the fraction  with N=4 ($p_4$) and N=6 ($p_6$) are $0.16$ and
$0.29$, respectively, which agrees well with results in
Ref.~\onlinecite{Patrykiejew:01lm}.

\section{Discussion and Conclusions}

In this work we have considered the recently developed crystal
phase field model \cite{Elder:04rq} in the presence of an external
pinning potential. As the model naturally incorporates both
elastic and plastic deformations, it provides a continuum
description of lattice systems such as adsorbed atomic layers or
2D vortex lattices, while still retaining the discrete lattice
symmetry of the solid phase. The main advantage of the model as
compared to traditional approaches is that despite retaining
spatial resolution on an atomic scale its temporal evolution
naturally follows diffusive time scales. Thus the numerical
simulations studies of the dynamics of the systems such as
approach to equilibrium can be achieved over realistic time scales
many order of magnitudes over the microscopic atomic models. We
have taken advantage of this and considered the full phase diagram
of the model as a function of the lattice mismatch and pinning
strength, both analytically and numerically. A systematic mode
expansion analysis has been used to determine the ground states
and the commensurate-incommensurate transitions in the model.
Numerical minimization of the corresponding free energy shows good
agreement with the analytical predictions and provides details on
the topological defects in the transition region. In particular,
we find that the transition remains discontinuous for all values
of the mismatch studied here. We have also performed a detailed
Voronoi analysis of the domain walls throughout the transition
region. Our results are consistent with simulations for atomistic
models of pinned overlayers on surfaces.

A particularly interesting application of the present model is to
pinned systems which are driven by external force. Examples of
such systems include driven adsorbed monolayers \cite{Robbins01},
driven charge density waves \cite{Sethna01} and driven flux
lattices \cite{Nori01}. Work in these problems is already in
progress.

\begin{acknowledgments}
This work has been supported in part by the Academy of Finland
through its Center of Excellence grant for the COMP CoE. M.K. has
been supported by NSERC of Canada. E.G. has been supported by
Funda\c c\~ao de Amparo \`a Pesquisa do Estado de S\~ao Paulo -
FAPESP (grant no. 03/00541-0).  K.R.E. acknowledges support from
the National Science Foundation under Grant No. DMR-0413062

\end{acknowledgments}

\bibliographystyle{apsrev}

\end{document}